# Geographical and Topology Control based Opportunistic Routing for Ad Hoc Networks

Ning Li, *Member IEEE*, Zhaoxin Zhang, Jose-Fernan Martinez-Ortega, Xin Yuan

*Abstract*—The opportunistic routing has great advantages on improving packet delivery probability ($P_{sc}$) between the source node and candidate forwarding set. For improving $P_{sc}$ and reducing energy consumption and network interference, in this paper, we propose an efficient and reliable transmission power control based opportunistic routing (ERTO) for wireless ad hoc networks. In ERTO, the network interference and fading, which are critical to routing performances but not taken into account in previous works, are considered during the prediction of $P_{sc}$. The $P_{sc}$, the expected energy consumption, and the relationship between transmission power and node degree are considered to optimize the optimal transmission power and forwarding node degree jointly. For improving routing effectiveness and reducing computational complexity, we introduce the multi-objective optimization and Pareto optimal into ERTO. During the routing process, nodes calculate the optimal transmission power and forwarding node degree according to the properties of the Pareto optimal solution set. Based on these innovations, the energy consumption, the transmission delay, and the throughout have been improved greatly.

## I. INTRODUCTION

In ad hoc network, the routing algorithm design is important because it guarantees reliable and efficient data transmission between source node and destination node. There are two different kinds of routing strategies: deterministic routing and opportunistic routing. The main disadvantage of deterministic routing is that it simply applies the operations and principles inherited from legacy routing solutions which were initially conceived for wired networks. So, it cannot adapt well on wireless applications [1]. Thus, opportunistic routing has been proposed. Compared with deterministic routing, it can improve network capability successfully, especially packet delivery probability [2-3] between the sender and candidate forwarding set (CFS) [1-3]. Many opportunistic routing algorithms have been proposed in recent years, such as [4-8].

However, there are some limitations on these previous works. *First*, in the previous works which improve packet delivery probability by power control, network interference or fading (which cannot be neglected in practical) is not taken into account. For instance, in [4], [5], and [7], both interference and fading channel are not considered; in [6] and [8], the interference between different nodes are ignored. *Second*, in previous works (such as [6] and [8]), one-hop energy consumption is not considered. Even in [4] and [5], the one-hop energy consumption is taken into account, the model used in these algorithms is not practical. For instance, during estimating one-hop energy consumption, the interference and fading are not used in these two algorithms, so the performance in practice is not as good as that in simulation. *Finally*, in previous works, the transmission power and relay node degree are optimized separately. However, considering the conclusions in [9], when nodes are uniformly distributed, the number of nodes in the coverage area under specific transmission power follows a specific distribution. So, during topology control, the transmission power and the number of nodes in CFS should be optimized jointly rather than separately. For instance, if the optimal transmission power is $P$ and the optimal forwarding node number is $n$ by mathematical calculation, but the combination of this transmission power and number of nodes in CFS may not exist or exist with low probability in practice, the routing performance will deteriorate.

For solving the limitations of prior arts, in this paper, we propose a probability prediction based effective and reliable topology control algorithm for opportunistic routing, shorted as ERTO. Our proposed algorithm can solve the limitations of prior arts. *First*, compared with previous works, our proposed algorithm does not need probe packets, so the cost of communication resources is reduced. *Second*, during the prediction of $P_{sc}$ and one-hop energy consumption, our proposed algorithm takes both fading and network interference into account. *Third*, note the fact that the transmission power and the number of forwarding nodes are related, so we optimize the transmission power and the number of forwarding nodes jointly to increase $P_{sc}$ and reduce one-hop energy consumption. *Finally*, for solving the NP-hard problem presented in Section I.B, we introduce the Pareto optimal solution set into this algorithm to achieve tradeoff between these three optimization objectives and reduce the computational complexity.

## II. NETWORK MODEL

Suppose the nodes in network are deployed uniformly [9]. Two nodes can communicate with each other only when the distance between these two nodes is smaller than their transmission ranges. For instance, as shown in the Fig. 1, the node $s$ and node $r$ can communicate with each other when $d_{sr} \leq r_s$ and $d_{sr} \leq r_r$, where $d_{sr}$ is the Euclidean distance between node $s$ and node $r$, $r_s$ and $r_r$ are the transmission ranges of node $s$ and node $r$, respectively. In this paper, we assume that the nodes in the network know their locations and can exchange their locations periodically. Only the neighbor nodes whose distances to destination node are smaller than source node have chance to be selected as candidate forwarding nodes. Therefore, we define the candidate forwarding area as follows.

**Definition 1:** The candidate forwarding area of node $s$ for transmitting data to destination node $d$ is defined as the intersection area of two circles $C(s, r_s)$ and $C(d, d_{ds})$, i.e., the red area in Fig.1; $C(d, d_{ds})$ means the circle with the center is destination node $d$ and radius is $d_{ds}$.

**Definition 2:** The CFS of node $s$ when transmits data packet to the destination node $d$ is defined as the set of nodes which locate in the candidate forwarding area, which can be expressed as: $R_{s \to d}(s, d_{ds}) = \{1, 2, \ldots, i | d_{di} \leq d_{ds}\}$.

**Definition 3:** The forwarding node degree of node $s$ is defined as the number of nodes whose distances to the destination node are smaller than node $s$, denoted by $n_{rel}$.

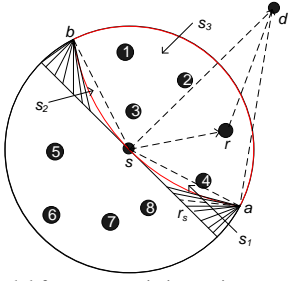

Fig. 1. The network model for opportunistic routing.

III. OPTIMIZATION MODEL

A. *Packet delivery probability between sender and CFS ($P_{sc}$)*

There are two different kinds of packet delivery probabilities: 1) the packet delivery probability between the sender and one forwarding node $i$ in CFS, known as $P_{si}$; 2) the packet delivery probability between the sender and CFS, known as $P_{sc}$. Notice that in opportunistic routing, the forwarding nodes in CFS are all one-hop neighbors of the sender. According to the definition of $P_{sc}$, which is the probability that the data packet sent by the *sender* can be received by *at least one forwarding node in CFS* successfully. The $P_{sc}$ can be calculated as:

$$P_{sc} = 1 - \prod_{i=1}^{n_{rel}}(1 - P_{si}) \quad (1)$$

where $n_{rel}$ is the forwarding node degree that is defined in the Definition 3. According to (1), we can conclude that $n_{rel}$ and $P_{si}$ can affect $P_{sc}$. So in the following of this section, we will investigate the effects of these two parameters on the performances of $P_{sc}$ in detail.

In the wireless network, whether the packet can be received successfully by the receiver relates to both the sender's transmission power and the interference at the receiver [10]. The interference at the receiver is defined as the summation of interference nodes' transmission power at the receiver [11]. The interference node is defined as the node whose transmission range covers this receiver. This is the natural properties of wireless communication. The nodes not only can affect their neighbors' data transmission but also can be affected by their interference nodes. Therefore, the $P_{si}$ shown in (1) is affected by both the transmission power of sender $s$ and the interference of receiver $i$. Moreover, not only interference but also fading has a great effect on $P_{si}$; however, calculating $P_{si}$ and $P_{sc}$ in opportunistic routing under the fading environment has not been investigated in previous works. Based on these facts, we propose the $P_{si}$ and $P_{sc}$ prediction algorithm for opportunistic routing under interference and fading environment.

**Corollary 1.** In interference and fading environment, the probability that the data packet sent by the sender can be received successfully by CFS is:

$$P_{sc}(P_{Ts}, n_{rel}) = 1 - \prod_{i=1}^{n_{rel}}(1 - exp\left[-\frac{\beta P_n d_{rs}^\eta}{P_{Ts} K}\right]\prod_{i=1}^{m}\frac{1}{1+\frac{\beta P_{Ti}}{GP_{Ts}(d_{ri}/d_{rs})^\eta}}) \quad (2)$$

where $P_{Ts}$ is the transmission power of sender $s$; $d_{sr}$ is the distance between sender and receiver; $\eta$ is the propagation loss coefficient, where $2 \leq \eta \leq 5$ depends on the geometry of propagation environment [12]; $K$ is the overall antenna gain and equals to $(G_t G_r \lambda^2)/((4\pi)^2 \Gamma)$, where $G_t$ and $G_r$ are the transmission and reception antenna gain, respectively; $\Gamma$ is the system loss; $\lambda$ is signal wavelength; $G$ is the processing gain [13]; $P_n$ is the noise power at receiver, $P_{Ti}$ is the transmission power of $ith$ interference nodes (except sender $s$), $d_{ri}$ is the distance between receiver and its interference node.

*Proof.* In the fading environment, given the transmission power of sender and CSMA/CA channel access, the received power at the receiver can be expressed as [14]:

$$P_{Rd} = \frac{K\alpha_{rs}^2}{d_{sr}^\eta}P_{Ts} \quad (3)$$

where $\alpha_{rs}^2$ is transmission exponential variable with unit mean due to Rayleigh fading [14].

Assuming that node $s$ wants to send packet to node $r$ and there are $m$ interference nodes around node $r$, and their distances to node $r$ are $d_{r1}$, $d_{r2}$, …, $d_{rm}$, respectively; the exponential variables characterizing the fluctuations in the powers of these nodes at node $d$ due to Rayleigh fading are $\alpha_{r1}^2$, $\alpha_{r2}^2$,….$\alpha_{rm}^2$; therefore, the interference and noise at node $d$ can be calculated as:

$$I_d = \frac{1}{G}\sum_{i=1}^{m}\frac{K\alpha_{ri}^2}{d_{ri}^\eta}P_{Ti} + P_n \quad (4)$$

Therefore, the signal interference noise ratio (SINR) $S_d(P_{Ts})$ at the receiver can be calculated as:

$$S_d(P_{Ts}) = \frac{P_{Rd}}{I_d} = \frac{\frac{K\alpha_s^2}{d_{rs}^\eta}P_{Ts}}{\frac{1}{G}\sum_{i=1}^{m}\frac{K\alpha_i^2}{d_{ri}^\eta}P_{Ti}+P_n} = \frac{\alpha_s^2}{\frac{1}{G}\sum_{i=1}^{m}\frac{\alpha_i^2}{(d_{ri}/d_{rs})^\eta}\frac{P_{Ti}}{P_{Ts}}+\frac{P_n d_{rs}^\eta}{P_{Ts}K}} \quad (5)$$

In the wireless networks, if the receiver can decode the received packet correctly, the SINR should above a certain threshold level $\beta$. According to (5) and the conclusions in [13], the predicted probability that SINR is above the given threshold in fading environment can be calculated as:

$$P(S_d \geq \beta) = P_{si} = P\left[\frac{\alpha_s^2}{\frac{1}{G}\sum_{i=1}^{m}\frac{\alpha_i^2}{(d_{ri}/d_{rs})^\eta}\frac{P_{Ti}}{P_{Ts}}+\frac{P_n d_{rs}^\eta}{P_{Ts}K}} \geq \beta\right]$$
$$= \int_0^\infty \int_0^\infty \cdots \int_0^\infty exp\left(-\beta\left[\frac{1}{G}\sum_{i=1}^{m}\frac{\alpha_i^2}{(d_{ri}/d_{rs})^\eta}\frac{P_{Ti}}{P_{Ts}}+\frac{P_n d_{rs}^\eta}{P_{Ts}K}\right]\right)\prod_{i=1}^{m}e^{-\alpha_i}d\alpha_i$$
$$= exp\left[-\frac{\beta P_n d_{rs}^\eta}{P_{Ts}K}\right]\prod_{i=1}^{m}\frac{1}{1+\frac{\beta P_{Ti}}{GP_{Ts}(d_{ri}/d_{rs})^\eta}} \quad (6)$$

Therefore, according to (1) and (6), the Corollary 1 holds. ∎

Note that, as shown in (2), the more nodes in CFS and the larger transmission power of the sender, the higher $P_{sc}$ is. However, in wireless ad hoc networks, on the one hand, the node energy is limited, so the smaller transmission power, the better performances; on the other hand, large transmission power causes much more serious interference, duplicate transmissions and heavier overhead than the small one. Therefore, the transmission power should be limited under the constraint of high packet delivery probability.

B. *Forwarding node degree*

As shown in (1), one parameter which can affect $P_{sc}$ is forwarding node degree. The forwarding node degree has been defined in Definition 3. Moreover, the number of nodes and transmission power are relevant [9]. When the nodes are uniformly distributed in the event area, the probability that the number of neighbors of node $s$ is $n$ can be expressed as [9]:

$$P(n) = \left(\frac{(\rho\Psi)^n}{n!}\right)e^{-\rho\Psi} \quad (7)$$

where $\Psi$ is the coverage area of node $s$ and can be calculated by $\Psi = \pi r_s^2$, $r_s$ is the transmission range of node $s$; $\rho$ is node density.

However, in ERTO, not the whole coverage area of the sender is taken into account during forwarding node selection. The interesting area is the candidate forwarding area which has been defined in Definition 1. So the (7) cannot be used directly to calculate the probability of forwarding node degree under specific transmission power. Therefore, we give the corollary 2 as follows.

**Corollary 2.** In ERTO, the candidate forwarding area $\Delta$ can be calculated as:

$$\Delta = r_s^2 \cdot arccos\frac{r_s}{2d_{ds}} + d_{ds}^2\left[\left(\pi - 2\,arccos\frac{r_s}{2d_{ds}}\right) - sin\left(\pi - 2\,arccos\frac{r_s}{2d_{ds}}\right)\right] \quad (8)$$

where the transmission range of sender $s$ is $r_s$ and the distance between sender $s$ and destination $d$ is $d_{ds}$.

*Proof.* This can be gotten easily based on the Fig. 1 and Geometry Theory. For saving space, we omit the proof. ∎

Therefore, the probability that there are $n_{rel}$ nodes in the candidate forwarding area when the transmission power is $P_{Ts}$ can be calculated as:

$$P_{rnd}(P_{Ts}, n_{rel}) = \left(\frac{(\rho\Delta)^{n_{rel}}}{n_{rel}!}\right) e^{-\rho\Delta} \quad (9)$$

The forwarding node degree and transmission power of the sender should make $P_{rnd}(P_{Ts}, n_{rel})$ as large as possible.

### C. Expected energy consumption

As discussed in Section 1, the larger transmission power, the higher packet delivery probability. However, when the transmission power is large, the energy consumption will be serious. This is not suitable for wireless nodes whose energy is limited. Therefore, in this section, we investigate the expected energy consumption model between the sender and CFS based on the conclusions in Section IV.A and Section IV.B. The conclusion is expressed by the Corollary 3.

**Corollary 3.** In ERTO, the one-hop expected energy cost that incurred by the node $s$ (with transmission power is $P_{Ts}$) sending a packet to the nodes in $R_{P_{Ts}}(s)$, can be calculated as:

$$\mathbb{C}_s(P_{Ts}, n_{rel}) = \frac{(n_{rel}E_r + \xi P_{Ts})\delta}{\left(1 - \prod_{i=1}^{n_{rel}}(1 - P_{si})\right)^2} \quad (10)$$

where $\delta = L/B$; $L$ is the data packet size and $B$ is bandwidth; $\xi$ is the consumption coefficient for data packet transmission; $P_{si}$ can be calculated based on (10), which has taken fading and interference into account; $E_r$ is the energy consumption for the reception.

*Proof.* Considering sender $s$ and its CFS $R_{P_{Ts}}(s)$ with transmission power $P_{Ts}$. Let $\mathbb{C}_s(P_{Ts}, n_{rel})$ denotes the one-hop expected energy cost incurred by node $s$ sends a packet to the nodes in $R_{P_{Ts}}(s)$. Therefore, according to [4] and [5], $\mathbb{C}_s(P_{Ts}, n_{rel})$ can be calculated as:

$$\mathbb{C}_s(P_{Ts}, n_{rel}) = \frac{E_s}{P_{sc}(P_{Ts}, n_{rel})} \quad (11)$$

where $E_s$ is the energy needed by the packet transmission and reception from the sender to CFS. Based on the conclusion in [4] and [5], the $E_s$ can be calculated as:

$$E_s = \left(|R_{P_{Ts}}(s)|E_r + E_{P_{Ts}}\right)E\{T_{P_{Ts}}(s)\} \cdot \frac{L}{B} \quad (12)$$

where $E_{P_{Ts}}$ is the energy consumption for data transmission with power $P_{Ts}$.

In this paper, for simplifying the calculation, we assume that the $L$ and $B$ keep constant during the calculation. $E\{T_{P_{Ts}}(s)\}$ is the mean energy consumption that the packet transmitted by the sender can be received by the receiver successfully by $T_{P_{Ts}}(s)$ times attempt, which can be calculated as [4]:

$$E\{T_{P_{Ts}}(s)\} = \sum_{l=0}^{\infty} l \times P_{tr}\{T_{P_{Ts}}(s) = l\} = \frac{1}{\left(1 - \prod_{i=1}^{n_{rel}}(1 - p_i)\right)} \quad (13)$$

where $P_{tr}\{T_{P_{Ts}}(s) = l\}$ is the probability that $T_{P_{Ts}}(s)$ equals to $l$ and can be calculated as: $P_{tr}\{T_{P_{Ts}}(s) = l\} = \left(\prod_{i=1}^{n_{rel}}(1-p_i)\right)^{l-1}(1-\prod_{i=1}^{n_{rel}}(1-p_i))$. Since $P_{sc}(P_{Ts}, n_{rel})$ has been calculated by (1), so the Corollary 3 holds. ∎

## IV. TOPOLOGY CONTROL BASED OPPORTUNISTIC ROUTING

### A. Optimal Solution Calculation

As shown in Section III.A, in opportunistic routing, the $P_{sc}$ relates to both forwarding node degree and $P_{Ts}$. Since the larger forwarding node degree is, the higher $P_{sc}$ is, the forwarding node degree should as large as possible. However, a large forwarding node degree needs large transmission power, which consumes more energy than the small transmission power. Unfortunately, on the one hand, the node energy is limited in wireless ad hoc networks; on the other hand, large transmission power causes much more serious interference than that in the small one; thus, the transmission power should be minimized as far as possible. These two optimal objectives are totally opposite and finding the optimal solutions for these issues has been proved a NP-hard problem [10]. This means that it is impossible to find an optimal solution that can make forwarding node degree and $P_{sc}$ maximum, while network interference and energy consumption minimal at the same time. Therefore, in this paper, we introduce the multi-objective optimization into the calculation of the optimal solutions to find the tradeoff between these optimal objectives. The issues introduced in Section I can be expressed as:

$$\begin{cases} \max(P_{sc}(P_{Ts}, n_{rel}), P_{rnd}(P_{Ts}, n_{rel})) \\ \min(\mathbb{C}_s(P_{Ts}, n_{rel})) \end{cases}$$
$$st. \ 0 \leq P_{Ts} \leq P_{max}$$
$$0 \leq n_{rel} \leq n \quad (14)$$

According to the multi-objective optimization theory [15-16], (14), it can be rewritten as:

$$\min_{x \in X^0} \mathbf{f}(\mathbf{x}) = (f_1(\mathbf{x}), f_2(\mathbf{x}), \dots, f_m(\mathbf{x})) \quad (15)$$

where $\mathbf{X}^0 = \{\mathbf{x} \in \mathbb{R}^n | g_i(\mathbf{x}) \geq 0, i = 1,2, \dots, p\}$ is the feasible region; $\mathbf{x} = (x_1, x_2, \dots, x_n)$ are the decision variables; $g_i(\mathbf{x}) \geq 0, i = 1,2, \dots, p$ are the constraint functions. In this paper, according to (14), we have: $\mathbf{x} = (P_{Ts}, n_{rel})$, $x_1 = P_{Ts}$, $x_2 = n_{rel}$; these functions represent the decision variables that we used in multi-objective optimization of (15); $\mathbf{f}(\mathbf{x}) = (-P_{sc}(P_{Ts}, n_{rel}), -P_{rnd}(P_{Ts}, n_{rel}), \mathbb{C}_s(P_{Ts}, n_{rel}))$, $f_1(\mathbf{x}) = -P_{sc}(P_{Ts}, n_{rel})$, $f_2(\mathbf{x}) = -P_{rnd}(P_{Ts}, n_{rel})$, $f_2(\mathbf{x}) = \mathbb{C}_s(P_{Ts}, n_{rel})$, these functions demonstrate the optimization objectives that will be calculated in (15) and they can be derived according to (2), (9), and (10), respectively; $g_1(\mathbf{x}) = P_{max} - P_{Ts}$, $g_2(\mathbf{x}) = n - n_{rel}$, these functions show the constraint functions of decision variables, i.e., the limitations of decision variables. So, the definition of Pareto optimal solutions is shown as follwos.

**Definition 4.** Assuming that $(\bar{P}_{TS}, \bar{n}_{rel}) \in \mathbf{X}^0$, if there is no $(P_{TS}, n_{rel}) \in \mathbf{X}^0$ which can make $f_i(P_{TS}, n_{rel}) < f_i(\bar{P}_{TS}, \bar{n}_{rel})$ ($i = 1,2,3$) true, then $(\bar{P}_{TS}, \bar{n}_{rel})$ is the Pareto Optimal Solution of (14); the set of all Pareto Optimal Solutions is Pareto Optimal Solution Set, denoted as $\mathbf{R}$.

The solutions in Pareto optimal solution do not mean that they can satisfy all the optimal objectives shown in (14) simultaneously. These optimal solutions can improve the performances of optimal objectives on at least one aspect compared with the solutions which are not in Pareto optimal solution set [15-16]. Moreover, the solutions in Pareto optimal solution set do not have comparability [15, 17-18]. For instance, solutions A and B are all in Pareto optimal solution set, then we cannot say solution A is better (or worse) than solution B.

There are two different kinds of approaches to calculate the Pareto optimal solution set of multi-objective optimization issues [15]: 1) the traditional approach; such as, method of objective weighting, method of distance functions, min-max formulation, etc.; the drawbacks of these traditional algorithms have been introduced in [15]; 2) the intelligent optimization approach; the intelligent optimization approach includes the genetic algorithm, the particle swarm optimization, etc. Since intelligent optimization approaches are much more accurate and effective than the traditional approach [16], in this paper, we use a genetic algorithm (GA) to calculate the Pareto optimal solution set of the issues shown in (14). For the GA based algorithms, the NSGA-II algorithm is the most popular and accurate one that used to calculate the Pareto optimal set. Moreover, the NSGA-II algorithm is simple and fast, which can reduce computational complexity further. The details of NSGA-II can be found in [16].

According to Definition 4, there are more than one solution in Pareto optimal solution set, so the Pareto optimal solution set $\mathbf{R}$ can be expressed as:

$$\mathbf{R} = \{(\bar{P}_{Ts1}, \bar{n}_{rel1}), (\bar{P}_{Ts2}, \bar{n}_{rel2}), \dots, (\bar{P}_{Tsn}, \bar{n}_{reln})\} \quad (16)$$

The solutions shown in (16) can be chosen as the optimal solution of (14). Notice that not all the optimal solutions in $\mathbf{R}$ can be chosen as the optimal solution of (14). This will be explained in the next section.

*B. Topology control algorithm*

In network, the nodes exchange information with their one-hop neighbors periodically; this information includes location and interference, which can be utilized to estimate $P_{sc}$ and expected energy consumption. The local information is updated based on the latest time-stamp in these periodic packets. When the nodes get the required information, they will calculate the Pareto optimal solution set. Based on the NSGA-II [16], the Pareto optimal solution set $\mathbf{R}$ can be got. According to the characteristics of the solutions in Pareto optimal solution set (which has been introduced in Section V.A), any solution in Pareto optimal solution set can be chosen as the final optimal solution. However, in practice, each node only has one transmission power and forwarding node degree, which means that not all the solutions in $\mathbf{R}$ can be selected. Therefore, according to the Pareto optimal solution set, current transmission power, and forwarding node degree, there are two different topology control strategies as follows.

1. The current transmission power and forwarding node degree are not in Pareto optimal solution set $\mathbf{R}$.

In this scenario, since the current transmission power and forwarding node degree are not in Pareto optimal solution set $\mathbf{R}$, the transmission power needs to be adjusted. As shown in [19], the relationship between transmission power and forwarding node degree is probabilistic in practice, i.e., when the transmission power is $P_{TS}^*$, the forwarding node degree is $n_{rel}^*$ with the probability $P_{rnd}(P_{TS}^*, n_{rel}^*)$, which is shown in (9) and Fig. 2. This means that the optimal solutions $(P_{TS}, n_{rel})$ in $\mathbf{R}$ may not exist or exist with low probability in reality. For instance, assuming that the current forwarding node degree is $n_{rel1}$, if the forwarding node degree needs to be reduced to $n_{rel2}$, there are many different values of $P_{TS}$ that can meet this requirement with different probabilities. Moreover, as shown in the Fig. 2, assuming that $(P_{Ts1}, n_{rel1}) \in \mathbf{R}$ can make $P_{sc}$ maximal and the expected energy consumption minimal at the same time, but the probability of $(P_{Ts1}, n_{rel1})$ calculated by (14) is very low, then if $P_{Ts1}$ is chosen as the optimal transmission power, it is highly probable that the forwarding node degree does not equal to $n_{rel1}$. So, the performances of $P_{sc}$ and expected energy consumption will deteriorate. Therefore, we choose the optimal solutions in $\mathbf{R}$ which can make (9) with the optimal value as the *candidate optimal solutions of* (14). However, according to the (9) and the Fig. 2, the $(P_{TS}, n_{rel})$ in $\mathbf{R}$ (which can make $P_{rnd}(P_{TS}, n_{rel})$ acquired the optimal value) is non-uniqueness.

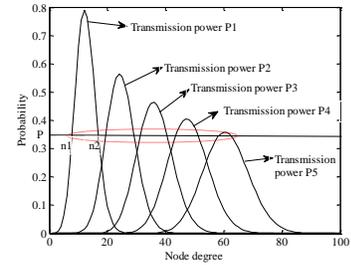

Fig. 2. The relationship between node degree and transmission power

As shown in the Fig. 2, if the optimal value of $P_{rnd}$ that calculated based on the solutions in Pareto optimal solution set is $P$, then the solutions $(P_{TS}, n_{rel})$ are not uniqueness; for instance, the solutions in the red area of the Fig. 2 have the same probability of $P$. So, the final the optimal transmission power during these solutions needs to be decided based on the values of $P_{sc}$ and expected energy consumption. Therefore, we define the optimal feasible solution set as follows.

**Definition 5.** When the optimal probability shown in (9) is $P$, which is calculated based on the solutions in Pareto optimal solution set, then the solutions which can make $P_{rnd}$ equal to $P$ are the elements of the optimal feasible solution set, denoted by $R^*$, where $R^* \in \mathbf{R}$.

For instance, as shown in the Fig.3, the solutions in the red area are all the elements of the optimal feasible solution set. According to the Definition 5, the feasible region has been reduced from Pareto optimal solution set to the optimal feasible solution set. The optimal transmission power and forwarding node degree will be chosen from the optimal feasible solution set $R^*$. Considering the fact that node performance will be decided by the worst node parameter, so for getting a balanced solution and reducing computational complexity, we propose the balanced optimal solution selection algorithm as follows.

In the balanced optimal solution selection algorithm, when the optimal solution is $(P_{Tsi}, n_{reli})$, then the corresponding $P_{sc}$, $P_{rnd}$, and $\mathbb{C}_s$ will be $P_{sc_i}(P_{Tsi}, n_{reli})$, $P_{rnd_i}(P_{Tsi}, n_{reli})$, and

$\mathbb{C}_{s_i}(P_{Tsi}, n_{reli})$, respectively. For different optimal solutions, these values are different. The optimal feasible solution set is $[(P_{Ts1}, n_{rel1}), (P_{Ts2}, n_{rel2}), \ldots, (P_{Tsm}, n_{relm})]$, and the corresponding performance set can be expressed as (since the values of $P_{rnd}$ are equal in optimal feasible solution set, so this set will not include it): $[(P_{sc_1}, \mathbb{C}_{s1}), (P_{sc_2}, \mathbb{C}_{s2}), \cdots, (P_{sc_m}, \mathbb{C}_{sm})]$, where $m$ is the number of solutions in optimal feasible solution set. The variance matrix of $P_{sc}$ and $\mathbb{C}_s$ can be expressed as: $v = [v_{P_{sc}}, v_{\mathbb{C}_s}]$. The parameters (i.e., $P_{sc}$ and $\mathbb{C}_s$) whose variance is larger will have a greater effect on the optimal solution selection than that of the smaller one [20-21].

Since the optimal solutions in the Pareto solution set do not have comparability and the values of $P_{rnd}$ are equal in the optimal feasible solution set, so the $P_{sc}$ and $\mathbb{C}_s$ shown in performance set have properties as follows.

**Corollary 4.** In optimal feasible solution set, if $P_{sc_i}(P_{Tsi}, n_{reli}) > P_{sc_j}(P_{Tsj}, n_{relj})$ or $P_{sc_i}(P_{Tsi}, n_{reli}) < P_{sc_j}(P_{Tsj}, n_{relj})$, there must exist $\mathbb{C}_{s_i}(P_{Tsi}, n_{reli}) > \mathbb{C}_{s_j}(P_{Tsj}, n_{relj})$ or $\mathbb{C}_{s_i}(P_{Tsi}, n_{reli}) < \mathbb{C}_{s_j}(P_{Tsj}, n_{relj})$, $i, j \in (1,2,3,\ldots,m)$, respectively; vice versa. Moreover, if $P_{sc_i}(P_{Tsi}, n_{reli}) = P_{sc_j}(P_{Tsj}, n_{relj})$, there must exist $\mathbb{C}_{s_i}(P_{Tsi}, n_{reli}) \neq \mathbb{C}_{s_j}(P_{Tsj}, n_{relj})$, vice versa.

*Proof.* As shown in (14), the purpose of multi objective optimization is to get large $P_{sc}$ and small $\mathbb{C}_s$ as far as possible. Assuming that $(P_{Ts1}, n_{rel1})$ and $(P_{Ts2}, n_{rel2})$ are in optimal feasible solution set $R^*$, then $P_{rnd_1}(P_{Ts1}, n_{rel1}) = P_{rnd_2}(P_{Ts2}, n_{rel2})$. Moreover, since $R^* \in \mathbf{R}$, so according to the Definition 5, to $P_{sc_1}(P_{Ts1}, n_{rel1})$, $\mathbb{C}_{s_1}(P_{Ts1}, n_{rel1})$, and $P_{rnd_1}(P_{Ts1}, n_{rel1})$, at least one of these parameters is better than that of $P_{sc_2}(P_{Ts2}, n_{rel2})$, $\mathbb{C}_{s_2}(P_{Ts2}, n_{rel2})$, and $P_{rnd_2}(P_{Ts2}, n_{rel2})$; meanwhile, at least one parameter of $P_{sc_1}(P_{Ts1}, n_{rel1})$, $\mathbb{C}_{s_1}(P_{Ts1}, n_{rel1})$, and $P_{rnd_1}(P_{Ts1}, n_{rel1})$ is worse than that of $P_{sc_2}(P_{Ts2}, n_{rel2})$, $\mathbb{C}_{s_2}(P_{Ts2}, n_{rel2})$, and $P_{rnd_2}(P_{Ts2}, n_{rel2})$. Since $P_{rnd_1}(P_{Ts1}, n_{rel1}) = P_{rnd_2}(P_{Ts2}, n_{rel2})$, so the Corollary 4 holds. ∎

**Corollary 5.** In optimal feasible solution set, the order of $P_{sc}$ from large to small is the same as that of the corresponding expected energy consumption, i.e., large $P_{sc}$ means large expected energy consumption.

*Proof.* The meaning of Corollary 5 can be explained as follows: in $R^*$, when $P_{sc}$ is the largest, then the corresponding $\mathbb{C}_s$ is the largest, too; when $P_{sc}$ is the second largest, then the corresponding $\mathbb{C}_s$ is the second largest, too; and so on. For proving the Corollary 5, we assume that the second largest $P_{sc}$ is $P_{sc_2}(P_{Ts2}, n_{rel2})$ and the corresponding expected energy consumption is $\mathbb{C}_{s_2}(P_{Ts2}, n_{rel2})$. If $\mathbb{C}_{s_2}(P_{Ts2}, n_{rel2})$ is not the second largest, then according to the properties of Pareto optimal solutions [15], there must exist $P_{sc_x}(P_{Tsx}, n_{relx})$ and $\mathbb{C}_{s_x}(P_{Tsx}, n_{relx})$ which satisfy $P_{sc_2}(P_{Ts2}, n_{rel2}) > P_{sc_x}(P_{Tsx}, n_{relx})$ and $\mathbb{C}_{s_2}(P_{Ts2}, n_{rel2}) < \mathbb{C}_{s_x}(P_{Tsx}, n_{relx})$; this conclusion does not conform the conclusion of the Corollary 4. So, the Corollary 5 is proved. ∎

The conclusion of Corollary 5 means that in $R^*$, when $P_{sc}$ is large, the corresponding $\mathbb{C}_s$ is large, too; vice versa. However, our purpose is to find an optimal solution that can make $P_{sc}$ the largest while $\mathbb{C}_s$ is the smallest, which is impossible in $R^*$ according to the conclusion of the Corollary 5. So, we need to find a tradeoff between $P_{sc}$ and $\mathbb{C}_s$. For getting the balanced optimal solution, based on the conclusion of the Corollary 5, we choose the intermediate value of $P_{sc}$ and corresponding $\mathbb{C}_s$ in optimal feasible solution set as the optimal solution, because this solution is more balanced than the other solutions in optimal feasible solution set. The intermediate value can be calculated based on the principles shown as follows:

(1) when the number of solutions in $R^*$ is odd, then the optimal solution is $(P_{Ts\frac{(m+1)}{2}}, n_{rel\frac{(m+1)}{2}})$, where $m$ is the number of solutions in $R^*$;

(2) when $m$ is even, two intermediate values can be gotten: $(P_{sc\frac{m}{2}}(P_{Ts\frac{m}{2}}, n_{rel\frac{m}{2}}), \mathbb{C}_{s\frac{m}{2}}(P_{Ts\frac{m}{2}}, n_{rel\frac{m}{2}}))$ and $(P_{sc\frac{(m+2)}{2}}(P_{Ts\frac{(m+2)}{2}}, n_{rel\frac{(m+2)}{2}}), \mathbb{C}_{s\frac{(m+2)}{2}}(P_{Ts\frac{(m+2)}{2}}, n_{rel\frac{(m+2)}{2}}))$. So, we need to choose the most appropriate one as the optimal solution. In this algorithm, the selection of optimal solution is based on the fact that the parameters whose variance are larger having greater effects on the algorithm's performances than the smaller one [20-21]; therefore, for these two intermediate values, if $v_{Psc} > v_{\mathbb{C}_s}$ and $P_{sc\frac{m}{2}}(P_{Ts\frac{m}{2}}, n_{rel\frac{m}{2}}) > P_{sc\frac{(m+2)}{2}}(P_{Ts\frac{(m+2)}{2}}, n_{rel\frac{(m+2)}{2}})$ (where $v_{Psc}$ means the variance of $P_{sc}$), $(P_{Ts\frac{m}{2}}, n_{rel\frac{m}{2}})$ will be chosen as the optimal solution; otherwise, if $v_{Psc} > v_{\mathbb{C}_s}$ and $P_{sc\frac{m}{2}}(P_{Ts\frac{m}{2}}, n_{rel\frac{m}{2}}) < P_{sc\frac{(m+2)}{2}}(P_{Ts\frac{(m+2)}{2}}, n_{rel\frac{(m+2)}{2}})$, $(P_{Ts\frac{(m+2)}{2}}, n_{rel\frac{(m+2)}{2}})$ will be chosen as the optimal solution. If $v_{Psc} < v_{\mathbb{C}_s}$ and $\mathbb{C}_{s\frac{m}{2}}(P_{Ts\frac{m}{2}}, n_{rel\frac{m}{2}}) > \mathbb{C}_{s\frac{(m+2)}{2}}(P_{Ts\frac{(m+2)}{2}}, n_{rel\frac{(m+2)}{2}})$, $(P_{Ts\frac{(m+2)}{2}}, n_{rel\frac{(m+2)}{2}})$ will be chosen as the optimal solution; otherwise, if $v_{Psc} > v_{\mathbb{C}_s}$ and $\mathbb{C}_{s\frac{m}{2}}(P_{Ts\frac{m}{2}}, n_{rel\frac{m}{2}}) < \mathbb{C}_{s\frac{(m+2)}{2}}(P_{Ts\frac{(m+2)}{2}}, n_{rel\frac{(m+2)}{2}})$, $(P_{Ts\frac{m}{2}}, n_{rel\frac{m}{2}})$ will be chosen as the optimal solution.

2. The current transmission power and forwarding node degree are in Pareto optimal solution set **R**.

In this situation, since the current transmission power and forwarding node degree are in Pareto optimal solution set **R**, which are better than the solutions that are not in Pareto solution set, for reducing the control cost, the nodes do not adjust their transmission power. The reasons that the nodes in this scenario do not adjust their transmission powers are: on the one hand, the network topology in ad hoc network changes frequently due to node mobility or node failure, so the frequent transmission power adjustment will consume a large amount of network resources that could have been used in packet transmission; on the other hand, the current transmission power and forwarding node degree in Parte optimal solution set **R** means that the current network performance is good. So, considering the energy consumption and control cost by controlling network topology, this tradeoff is worthy.

### C. Topology control based opportunistic routing algorithm

When the optimal transmission power and forwarding node degree have been gotten, the sender adjusts its transmission power and transmits a data packet to the candidate forwarding set. For achieving high $P_{sc}$, all the nodes in the candidate forwarding area will be chosen as forwarding nodes. Because the $P_{si}$ of each forwarding node $i$ has been calculated during topology control, so for reducing the computational costs, in ERTO, we use except transmission account (ETX) as the performance metric to determine the priority of each forwarding node in the candidate set. The ETX used in this paper is similar but different from the traditional ETX which is defined in ExOR [22]. Because in this paper, the calculation of ETX takes both network interference and fading into account. Therefore, the ETX can be calculated as:

$$ETX(i) = 1/P_{si} \quad (17)$$

where $P_{si}$ can be calculated by (6).

In the candidate set, the node which has small ETX has a high priority for packet transmission. Similar to that shown in ExOR, this process continues until the packet is received by the destination node.

**Corollary 6.** The computational complexity of ERTO by using NSGA-II is $O(MN^2)$, where $M$ is the number of optimization objectives and $N$ is the population size.

*Proof.* As introduced in Section IV.A, the main computational complexity of ERTO is the calculation of Pareto optimal solutions. In ERTO, the Pareto optimal solution is calculated by NSGA-II [16], in which the computational complexity is $O(MN^2)$, so the computational complexity of ERTO is $O(MN^2)$, where $M$ is the number of optimization objectives and $N$ is the population size. ∎

## V. SIMULATION AND DISCUSSION

The constant bit rate (CBR) [2, 23-24] is used in this simulation to generate data packet; each CBR data packet is transmitted between two nodes which are chosen randomly. The number of CBR connection pairs represents the traffic load in the network. The larger number of CBR connection pairs, the higher traffic load is. The simulation parameters can be found in the Table 1. For comparing the performances of ERTO, three opportunistic routing algorithms are implemented in this simulation: 1) ExOR [22]; 2) TCOR [4]; 3) EEOR [5].

TABLE 1. SIMULATION CONFIGURATION

| Parameter | Value |
|---|---|
| deployment area | $1000m \times 1000m$ |
| initial node transmission range | $200m$ |
| packet length | 1024bits |
| data rate | 15Kbps |
| initial energy | 5J |
| maximum transmission power | 0.8W |
| minimum transmission power | 0.1W |
| number of nodes | 40-120 |
| receiving power | 0.05W |
| number of CBR pairs | 20-100 |

### A. Performance under different number of nodes

The Fig. 3 illustrates the packet delivery probability of these four opportunistic routing algorithms. In the Fig. 3, the packet delivery probability of ERTO is much higher than that of the other three algorithms, such as 40% higher than ExOR and 20% higher than EEOR when the number of nodes is 100. With the increasement of the number of nodes, the packet delivery probability increases obviously in ExOR and ERTO, while the growth is slight in TCOR and EEOR. Moreover, when the number of nodes is large, the increase in ExOR is smaller than that in ERTO. Two parameters can explain this conclusion: the node degree and network interference. When the network density is small, the node degree is the domain parameter on determining packet delivery probability; when the node degree increases, the packet delivery probability will also increase. However, with the rise of the node density, the network interference becomes more and more serious than that in the sparse network; then the network interference will be the domain parameter. Since ERTO takes network interference into account, the packet delivery probability of ERTO increases even in the dense network.

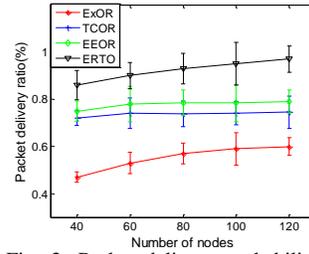
Fig. 3. Packet delivery probability under different node densities

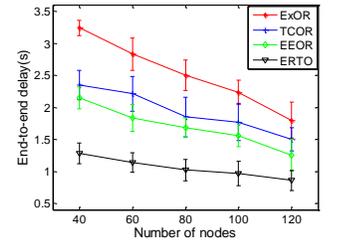
Fig. 4. End-to-end delay under different node densities

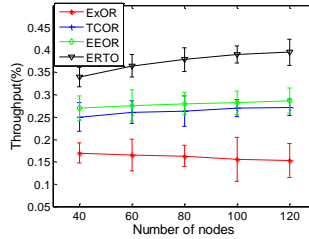
Fig. 5. Throughput under different node densities

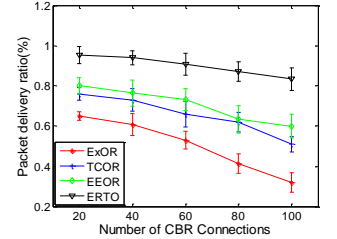
Fig. 6. Packet delivery probability under different traffic load

The performances of end-to-end delay is presented in Fig. 4. On the one hand, high packet delivery probability means low probability of packet retransmission, which reduces transmission delay; so with the increasement of node density, the transmission delay decreases. On the other hand, the number of transmission hops can also affect the performances of end-to-end delay. Large hop numbers mean long transmission delay. Therefore, considering both the packet delivery probability and the number of transmission hops, the decrease of transmission delay in ExOR is larger than that in the other three algorithms. Moreover, since the packet delivery probability in ERTO is better than that in ExOR, EEOR, and TCOR, the transmission delay in ERTO is the smallest.

In Fig. 5. The throughput of ERTO is the highest among these four algorithms. With the increasement of network density, the rise of throughput is slight in these four algorithms. The throughput of ERTO grows faster with the increasement of node density than the other three algorithms. This can be explained as follows. On the one hand, when the network density rises, the packet delivery probability also increases, so the throughput increases with the increasement of the node density; on the other hand, the more nodes in the network, the more serious interference, which deteriorates the performances of throughput.

### B. Performance under different traffic load

In Fig. 6, with the increase of traffic load, both the packet delivery probabilities of these four algorithms decrease. Moreover, the decrease ratio is the largest in ExOR, while it is the smallest in ERTO. Similar results can also be found in Fig. 7. In Fig. 7, when traffic load increases, on the one hand, the network congestion and contention increase, so transmission delay increases; on the other hand, due to the packet delivery probability decreases, the transmission delay increases further. However, since ERTO can adjust transmission power during the routing process based on the network interference, the performances of packet delivery probability and transmission delay in ERTO are all better than that in the other three algorithms.

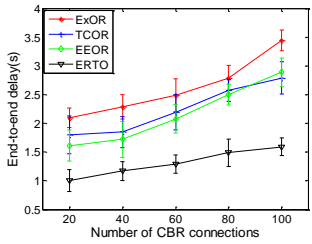

Fig. 7. End-to-end delay under different traffic load

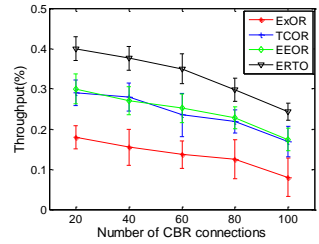

Fig. 8. Throughput under different traffic load

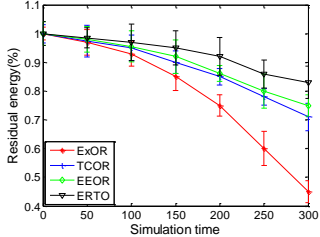

Fig. 9. Residual energy of ExOR, TCOR, EEOR, and ERTO

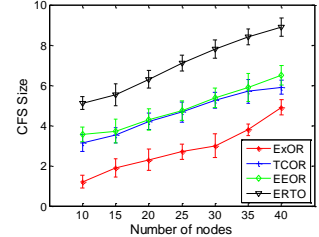

Fig. 10. CFS sizes under low network densities

The throughput of these four algorithms under different traffic load has been shown in Fig. 8. In Fig. 8, the performances of throughput in ERTO is also the best. When the traffic load increases, the throughput decreases. These can be explained as: 1) similar to the performances of packet delivery probability and transmission delay, when the traffic load increases, the network congestion and contention increases, which deteriorates the performances of network throughput; 2) when the traffic load increases, both packet delivery probability and transmission delay become worse than that when the traffic load is small, so the throughput decreases further.

### C. Performances of energy consumption under different simulation time

The energy consumption of ERTO is less than that of the other three algorithms, which is present in Fig. 9. In Fig. 9, with the increasement of simulation time, the residual energy of these four algorithms reduces. However, the reduction of ERTO is smaller than that of ExOR, EEOR, and TCOR. Even EEOR, TCOR, and ERTO take energy consumption into account during the routing process, considering the performances of packet delivery probability and transmission delay, the energy consumption in ERTO is less than that in EEOR and TCOR. In Fig. 10, the CFS sizes in different routing algorithms under different network densities are presented. With the increasement of the network density, the CFS sizes in these four routing algorithms also increase. If network density is large enough, the CFS sizes of ERTO, TCOR, and EEOR become stable. However, the CFS size of ExOR will keep increasing. This is because ERTO, TCOR, and EEOR can control transmission power, while ExOR cannot. Moreover, since the packet delivery probability of ERTO is higher than that of the other three algorithms, the CFS size of ERTO is also larger than the other algorithms.